\documentclass[11pt]{emulateapj}
\usepackage{graphicx}
\usepackage{multirow}
\usepackage{textcomp}
\usepackage{epsfig}
\usepackage{amsmath}
\usepackage{amssymb}
\usepackage{amsthm}
\usepackage{xy}

\def\lowba{SGR\,0418$+$5729}

\def\3xmm{3XMM\,J1852$+$0033}
\def\lowbb{Swift\,J1822.3$-$1606}

\newcommand{\xmm}{{\em XMM--Newton}}
\newcommand{\XMM}{{\em XMM--Newton}}

\def\ergs {erg\,s$^{-1}$}
\def\ergscm2 {erg\,s$^{-1}$cm$^{-2}$}
\def\ss {s\,s$^{-1}$}
\def\cm2 {cm$^{-2}$}
\def\arcsec{$^{\prime\prime}$}

\shortauthors{N. Rea et al.}

\begin{document}

\title{3XMM J185246.6$+$003317: another low magnetic field magnetar}

\author{N. Rea\altaffilmark{1,2},  D. Vigan\`o\altaffilmark{1}, G. L. Israel\altaffilmark{3}, J. A. Pons\altaffilmark{4}, D. F. Torres\altaffilmark{1,5}}

\altaffiltext{1}{Institute of Space Sciences (CSIC--IEEC), Campus UAB,  Torre C5, 2a planta, 08193 Barcelona, Spain.}
\altaffiltext{2}{Astronomical Institute "Anton Pannekoek", University of Amsterdam, Postbus 94249, NL-1090 GE Amsterdam, the Netherlands.}           
\altaffiltext{3}{INAF-Astronomical Observatory of Rome, via Frascati 33, 00040, Monte Porzio Catone, Roma, Italy.}            
\altaffiltext{4}{Departament de Fisica Aplicada, Universitat d'Alacant, Ap. Correus 99, 03080 Alacant, Spain.} 
\altaffiltext{5}{Instituci\'o Catalana de Recerca i Estudis Avan\c{c}ats (ICREA), Barcelona, Spain.}

\begin{abstract}

We study the outburst of the newly discovered X-ray transient  3XMM J185246.6$+$003317, re-analysing all available \XMM\, observations of the source to perform a phase-coherent timing analysis, and derive updated values of the period and period derivative. We find the source rotating at $P=11.55871346(6)$\,s (90\% confidence level; at epoch MJD 54728.7) but no evidence for a period derivative in the 7 months of outburst decay spanned by the observations. This translates in a 3$\sigma$ upper limit for the period derivative of $\dot{P}<1.4\times10^{-13}$\ss , which, assuming the classical magneto-dipolar braking model, gives a limit on the dipolar magnetic field of $B_{\rm dip}<4.1\times10^{13}$\,G . The X-ray outburst and spectral characteristics of 3XMM J185246.6$+$003317 confirms the identification as a magnetar, but the magnetic field upper limit we derive defines it as the third "low-B" magnetar discovered in the past three years, after \lowba\, and \lowbb.  We have also obtained an upper limit to the quiescent luminosity ($< 4 \times 10^{33}$ \ergs), in line with the expectations for an old magnetar. The discovery of this new low field magnetar reaffirms the prediction of about one outburst per year from the hidden population of aged magnetars. 

\end{abstract}


\section{Introduction}

Neutron stars are the relic of the supernova explosions of massive stars \citep{baade34}.  Five decades after their discovery \citep{hewish68}, these compact objects have appeared in many different flavors. The most common are radio pulsars, usually modeled as rapidly rotating magnetic dipoles. 
Another important sub-group is formed by binary neutron stars, either as X-ray pulsars accreting from a companion or normal radio pulsars orbiting a companion star. Perhaps the most intriguing class among isolated neutron stars are the "magnetars", so called because they are believed to be powered by their super strong magnetic field (see \citealt{mereghetti08,rea11} for recent reviews). Initially classified as Anomalous X-ray Pulsars (AXPs) and Soft Gamma Repeater (SGRs), it is now accepted that this is not an intrinsic distinction, but rather a historical nomenclature due to the different ways they were discovered: as a steady emitter visible in X-ray surveys, or during a high energy burst or flare from a new direction in the sky. Magnetars are characterized by rotational periods in the 0.3--12\,s range, period derivatives between $10^{-15}-10^{-10}$\ss , X-ray luminosities of $10^{31}-10^{35}$\ergs, and episodes of enhanced X-ray persistent emission either as a long-lived radiative outburst (lasting months-years) or short bursts and flares (lasting seconds-minutes). Both their steady and transient X-ray phenomena are powered by their strong magnetic fields, that can stress the neutron star crust causing stellar quakes, accompanied by global magnetospheric reorganizations with the subsequent powerful high energy emission. Eventually, shorter flares might instead be purely magnetospheric, caused by re-connection of magnetic field lines higher up in the magnetosphere \citep{tlk02,lyutikov03}.

In 2009, a peculiar magnetar \citep{vanderhorst10,esposito10,rea10} was discovered during an active epoch (\lowba ), as many other members of the magnetar class, but in this case its estimated surface dipolar magnetic field (at the equator) of $B = 6.2\times10^{12}$\,G \citep{rea13} was rather low, more typical of a normal radio pulsar. Some years later, another ``low magnetic field magnetar'' was discovered (\lowbb\,: $B\sim2\times10^{13}$\,G; \citealt{rea12,scholz12}), again showing all the characteristics of the outburst activity of a typical magnetar. A plausible solution to the apparent puzzle considers these objects  as aged magnetars, that have largely dissipated their external dipolar field, but still holding a crustal/internal field one or two orders of magnitude larger. This internal field would be responsible for the bursting activity and intense outbursts \citep{rea10,turolla11}. This scenario has been strengthened by detailed studies of the evolution of neutron stars endowed with strong magnetic fields, and applied to the two known low field magnetars  \citep{pons09,vigano13,rea13}.
Furthermore, the absorption feature observed during the outburst of the lowest field magnetar, \lowba, if interpreted as a proton cyclotron feature, confirms a $\sim10^{14}$\,G magnetic field in a magnetic loop close to the surface \citep{tiengo13}.

In this Letter we have re-analized all the archival \XMM\, observation of 3XMM J185246.6$+$003317 (hereafter \3xmm): a new transient source discovered serendipitously while undergoing an outburst in 2008 \citep{zhou13}. We first report on the data analysis, and in \S\ref{discussion} we argue that this source is a low magnetic field magnetar, discussing the consequences of this finding in terms of the population of old magnetars and their magneto-thermal evolutionary path.

\begin{table}
\caption{Summary of the {\it XMM-Newton} observations.}
\label{obslog}
\footnotesize{
\begin{tabular}{@{}lcccc}
\hline
\hline
Obs. ID  &  Obs. Date & Camera & Exposure &  count rate  \\ 
& YYYY-MM-DD & & (ks) & (count/s) \\
\hline
 Quiescence & & & & \\
 \hline
0204970201 & 2004 Oct 18 &  MOS 2 & 31.4 & $<$0.004 \\
	&  & MOS 1 &  31.4 &  $<$0.005 \\
0204970301 & 2004 Oct 24 &  MOS 2 & 31.1 & $<$0.004\\
	&  & MOS 1 &  31.4 &  $<$0.005 \\
0400390201 & 2006 Oct 08 & MOS2 & 30.4 & $<$0.005 \\	
	0400390201 & 2007 Mar  20 & MOS2 & 34.4 & $<$0.007  \\
		&  & MOS 1 &  34.4 &  $<$0.006\\	
		\hline
Outburst&  & &  & \\
\hline
0550670201 & 2008 Sep 19 &  MOS 2 & 21.6 & 0.210$\pm$0.003\\
0550670301 & 2008 Sep 21 & MOS 2 & 30.3 & 0.199$\pm$0.003  \\ 
0550670401 & 2008 Sep 23 & MOS 2 & 35.4 & 0.196$\pm$0.003 \\
0550670501& 2008 Sep 29 & MOS 2 & 33.3 &  0.198$\pm$0.003\\
0550670601 & 2008 Oct 10 & MOS 2 & 35.5 & 0.148$\pm$0.002 \\
0550671001 & 2009 Mar 16 & MOS 2 & 27.2 & 0.033$\pm$0.001 \\
					&  & MOS 1 &  27.2 & 0.033$\pm$0.001 \\
0550670901& 2009 Mar 17 & MOS 2 & 26.2 & 0.030$\pm$0.001  \\
					&  & MOS 1 &  26.2 & 0.033$\pm$0.001\\
0550671201 & 2009 Mar 23 & MOS 2 & 27.1 &  0.030$\pm$0.001\\
					&  & MOS 1 &  27.1 & 0.032$\pm$0.001\\
0550671101 & 2009 Mar 25 & MOS 2 & 18.8 & 0.028$\pm$0.002 \\
					&  & MOS 1 & 19.6 & 0.033$\pm$0.001 \\
0550671301  & 2009 Apr 04 & MOS 2 & 26.2 & 0.013$\pm$0.007 \\
					&  & MOS 1 & 26.2 & 0.028$\pm$0.001  \\
0550671901 & 2009 Apr 10 & MOS 2 & 30.7 & 0.023$\pm$0.001 \\
					&  & MOS 1 &  30.6 & 0.024$\pm$0.001 \\ 
0550671801 & 2009 Apr 22 & MOS 2 & 28.2 &   0.026$\pm$0.001  \\
					&  & MOS 1 &  28.2 &  0.024$\pm$0.001\\
\hline
\hline
\end{tabular}}
\end{table}


\begin{figure}
\centering
\vbox{
\includegraphics[width=9cm]{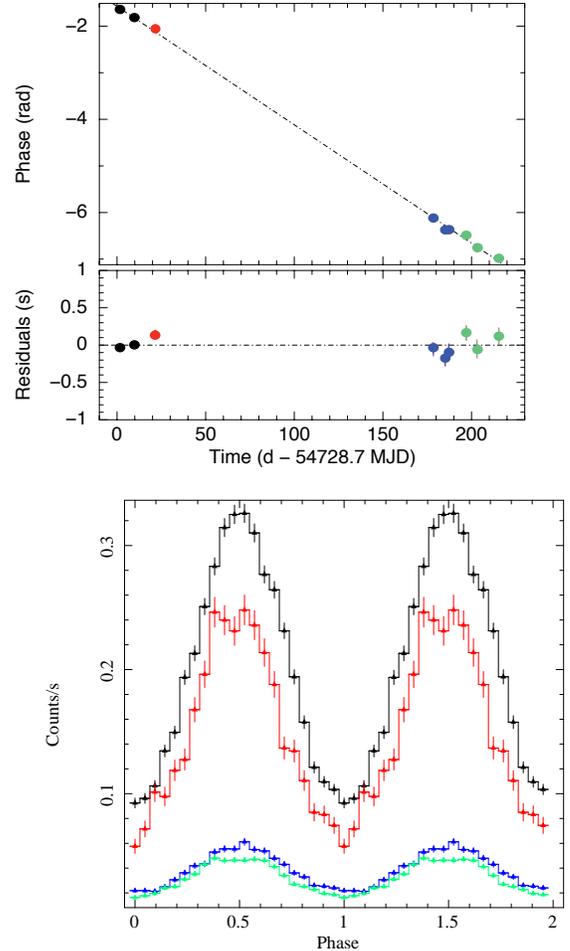}
\includegraphics[width=6cm,angle=270]{fig1b.eps}}
\caption{{\em Left}: \3xmm's pulse phases derived by fitting with a sine function the pulse profile folded with a trial period (see text for details, and below for the color code definition). The phase evolution in time is fitted with a linear function. The residuals with respect to our best phase-coherent solution are
reported in the lower panel, in units of seconds. {\em Right}: Pulse profiles in the 0.3--10\,keV energy range. From top to bottom they refer to: i) black - MOS2 observations performed between 2008 September 19--29, with a 0.5--10\,keV observed flux of $\sim4.2\times10^{-12}$ \ergscm2 (see also Figure\,\ref{fig:spectra}) , ii) red - 2008 October 10 (MOS2) at a flux of  $\sim2.8\times10^{-12}$\ergscm2 , iii) blue - 2009 March 16--25 (MOS1 and MOS2) at  $\sim6\times10^{-13}$\ergscm2 , and iv) green - 2009 April 4--22 (MOS1 and MOS2) at
$\sim5\times10^{-13}$\ergscm2 (note that some observations have been merged to increase the accuracy in the phase determination).}
\label{fig:timing}
\end{figure}



\begin{figure}[t]
\includegraphics[width=6.8cm,angle=270]{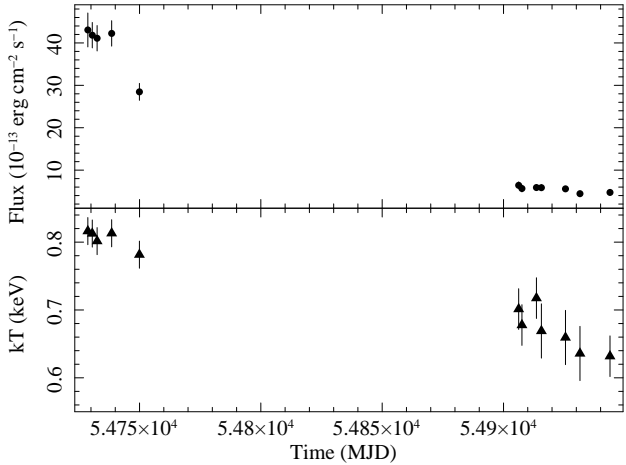}
\caption{Evolution of the 0.5--10\,keV observed flux and blackbody temperature as a function of time (see text for details).}
\label{fig:spectra}
\end{figure} 



\begin{figure*}
\centering
\includegraphics[width=8cm]{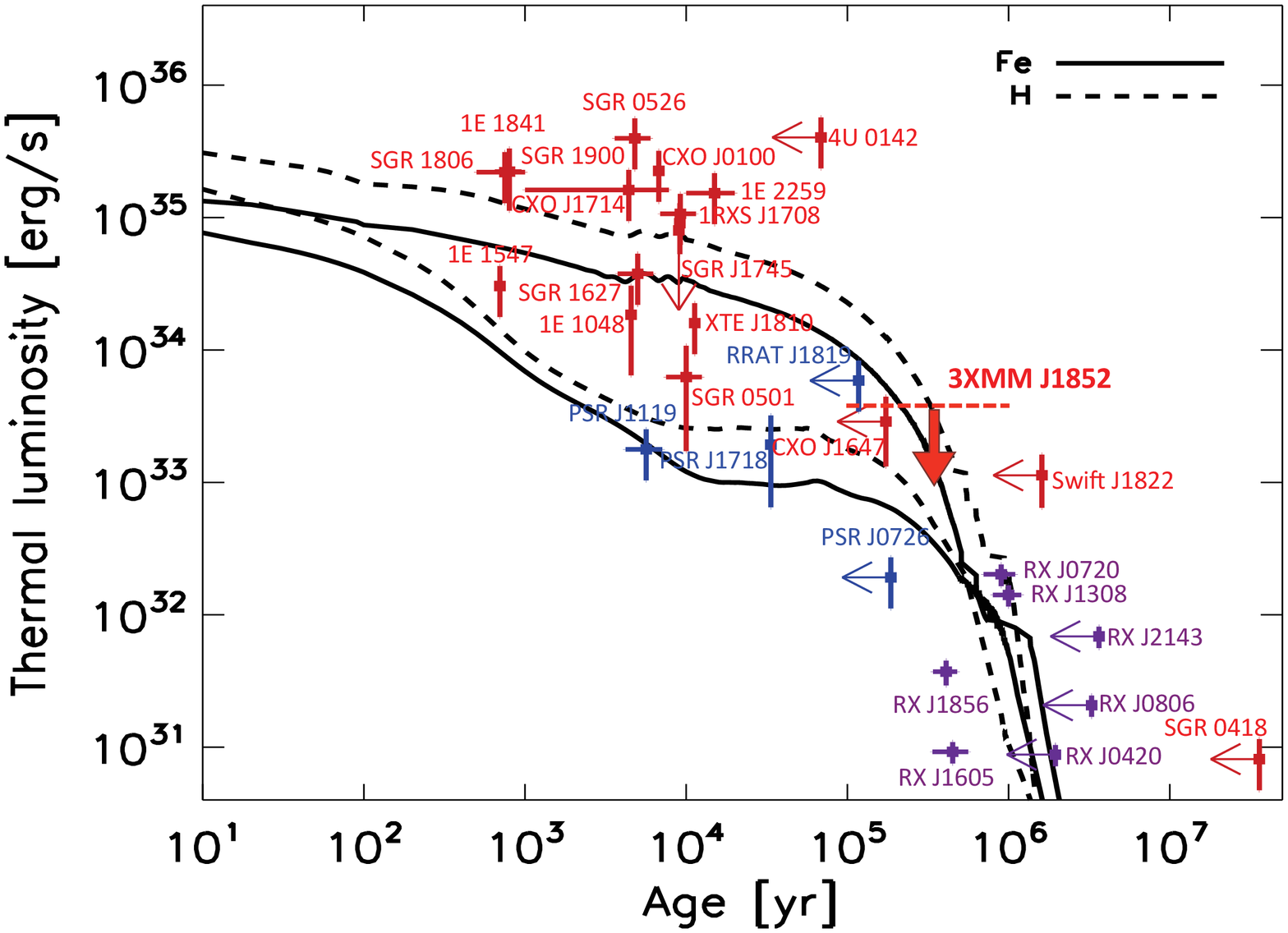}
\hspace{0.5cm}
\includegraphics[width=8cm]{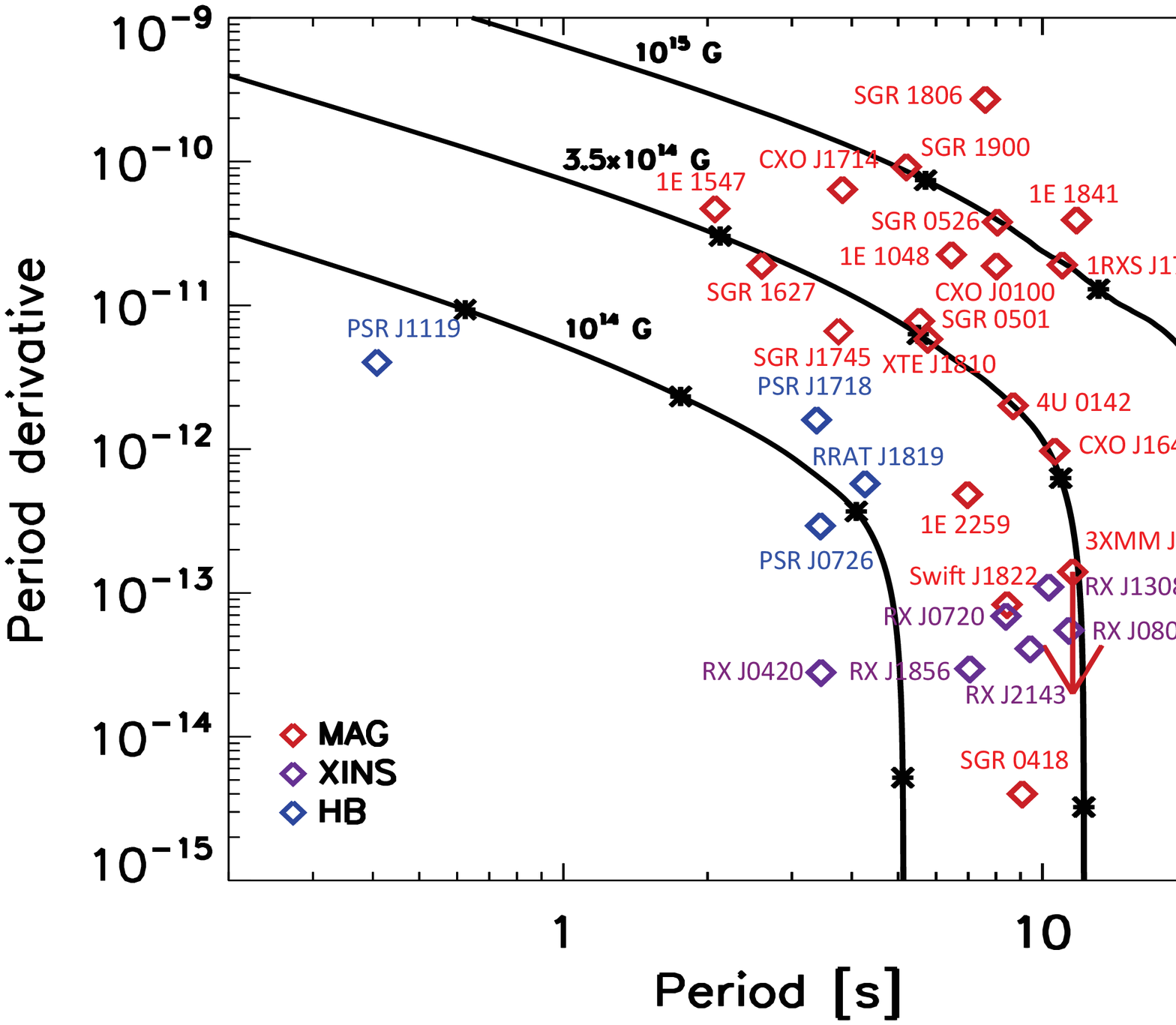}
\caption{\footnotesize{Comparison between observed properties of magnetars (red), high B pulsars (blue) and XINSs (purple), and the predictions from magneto-thermal models (lines, see \citealt{vigano13} for details about both observed and predicted values). {\em Left}: quiescent bolometric thermal luminosity versus age for an isolated neutron star born with a dipolar field (at the pole) of $B=10^{14}$ (top lines) and $B=10^{15}$ G (lower lines), for an iron (solid) or hydrogen envelope (dashed). Arrows indicate sources with no available alternative age estimate (e.g., SNR age) and large characteristic ages, which very likely overestimates the real age due to magnetic field dissipation. The limit we derive \3xmm 's quiescent luminosity is indicated with a red arrow, and the dashed line represents the limits on the magneto-thermal age (see text for details). {\em Right}: evolution of the timing properties for $B=10^{14}$, $B=3.5\times 10^{14}$ G and $B=10^{15}$ G, assuming an aligned rotator in the spin-down formula by \cite{spitkovsky06}. Asterisks indicate the real ages of $1, 10, 100$ and 500 kyr in descending order along the magnetic isotracks.}}
\label{fig:mt}
\end{figure*}


\section{XMM-Newton data analysis}
\label{analysis}

\3xmm\, was observed several times with \xmm\ \citep{jansen01}. Data have been processed using SAS version 13, and we have employed the most updated calibration files available at the time the reduction was performed (November 2013). The source was detected serendipitously only in the MOS cameras (\citealt{turner01}; see Table\,\ref{obslog}) in the 2008 and 2009 observations. The MOS1 and MOS2 cameras were set-up in Full Frame mode, with a timing resolution of 2.6\,s. We have applied standard data screening criteria in the extraction of scientific products.  Source photons were extracted from a circular region with a radius of 40\arcsec , and a similar circle was used for the background, in the same CCD of the source. We used the same extraction region to estimate the count rate upper limit for the four observations of the source in quiescence (see Table~\ref{obslog}). Our spectral analysis was restricted to photons having PATTERN$\leq$12 and FLAG=0. All photon arrival times have been referred to the Solar System barycenter (TDB time system and DE200 ephemeris).

\section{Outburst}
\subsection{Timing analysis}
\label{timing}

Timing analysis was performed using the  phase-fitting technique 
(details on this technique can be found in \citealt{dallosso03}), and using all ``outburst" data listed in Table~\ref{obslog} (both from the MOS1 and MOS2 cameras when available).  
We merged the photon arrival times of some contiguous pointings (2008 September 19-29, 2009 March 16-25, and 2009 April 4-22) in order 
to increase the phase accuracy and/or to reduce the datapoint scatter (see Figure\,\ref{fig:timing}).  Data were folded using a trial period of
11.5587072\,s at epoch 54728.7\,MJD. The phase of the modulation was inferred by fitting the average pulse shape  of each observation (folded with the above trial period), with one or more harmonics (the exact number was determined by requesting that the inclusion of any higher harmonic was statistically significant).  In almost all 
cases the use of the fundamental harmonic alone was sufficiently accurate. In Figure\,\ref{fig:timing} we plot the phases at which the fundamental sine function fitted to the pulse profile is equal to zero.

The time evolution of the phase can be described by a relation: $\phi=\phi_0+2\pi(t-t_0)/P-\pi(t-t_0)^2\dot{P}/P^2$. A linear fit of the resulting pulse phases, by assuming the initial trial period reported above, gives a reduced $\chi^2_r \sim 2$ 
for 8 degree of freedom (d.o.f. hereafter). The inclusion of a quadratic term in the phase modelling,  
corresponding to a first period derivative component, was not 
significant in our data with a 3$\sigma$ (2 parameters of interest, p.o.i) 
upper limit on the period derivative of $\dot{P} < 1.4\times 10^{-13}$\,s\,s$^{-1}$ 
with a reduced $\chi^2_r \sim 2.2$ (for 7 d.o.f.; see also Figure\,\ref{fig:timing}).  
The resulting best-fit solution corresponds to a spin period of $P=11.55871346(6)$\,s 
(90\% confidence level and 1 p.o.i; epoch 54728.7\,MJD). The new timing solution implies a r.m.s. 
variability of $<0.2$\,s .

The pulse profile seems rather stable in shape, a single peak remaining in phase during the outburst decay. However, we note that we caught the outburst at a late time,  
and a pulse profile stabilization was already observed in other low-B magnetars during the outburst decay. 
The pulsed fraction is relatively stable in time with an average value of $\sim62$\% (defined as the semi-amplitude of the fundamental sinusoidal modulation divided by the mean source count rate).

\subsection{Spectral analysis}
\label{spectral}

We have performed the spectral analysis using all the observations during the outburst phase of \3xmm\, reported in Table~\ref{obslog}, and data from both the MOS1 and MOS2 cameras when available. Spectra from the 2008 observations were grouped to have at least 50 counts per bin, while for the 2009 observations we
required at least 30 counts per bin. The rebinning was made with special care in order not to oversample the instrument spectral resolution by more than a factor of three. We used XSPEC 12.7.1 for the spectral fitting. Our results are consistent within errors with those of \cite{zhou13}. We report in Figure\,\ref{fig:spectra} the evolution of the spectral parameters and of the 0.5--10\,keV observed flux for a {\tt phabs}$*${\tt bbodyrad} spectral model ($\chi^2=1.1$ (1049 degrees of freedom); $N_{\rm H}=1.32(5)\times10^{22}$\cm2 ). Similarly to \cite{zhou13}, we find a tiny excess flux at energies higher than 6\,keV which is not properly modelled by a single blackbody model alone (although this is not influencing the goodness of the $\chi^2$). Using a Resonant Cyclotron Scattering model (RCS; \citealt{rea08}) this excess is instead well accounted for (note, however, that the RCS has two additional free parameters) by the non-thermal scattering tail. We refer to \citealt{zhou13} for other details about the spectral parameters inferred by this model.  Assuming a distance of 7.1\,kpc, we find a blackbody radius decreasing from 0.8\,km to 0.4\,km .

\section{Quiescence}

In order to derive a stringent upper limit to the flux of \3xmm\, during its quiescent state, we have extracted all photons encircled in a 40\arcsec\, radius around the position of the source, for the four available observations (Table~\ref{obslog}), and both the MOS1 and MOS2 cameras when available. 
We have added all the event files of the MOS1 and MOS2, created an image of the resulting merged event file, and used the {\tt Ximage} {\em sosta} tool to derive a source upper limit taking into account its PSF correction for the off-axis position, the vignetting and the sampling deadtime. This method uses the Bayesian approach with the prior function set to the prescription described in \cite{kraft91}. We have derived, for a total exposure time of 251\,ks, a 3$\sigma$ upper limit on the source count rate of 0.0014 count/s in the 0.3--10\,keV range. With this upper limit, assuming a distance of 7.1\,kpc, a 10\,km radius surface emission, and an $N_H=1.32\times10^{22}$\cm2 (as derived in \S \ref{spectral}), we can obtain the upper limits on the surface temperature and on the bolometric thermal luminosity during quiescence of kT$<$0.15\,keV, and $L_{qui}< 4\times10^{33}$\ergs .

\section{Discussion}
\label{discussion}

We have reported a phase coherent timing solution for the newly discovered transient \3xmm\, \citep{zhou13}, which underwent an outburst in 2008, and was caught serendipitously by \XMM\, during a series of observations of the SNR Kes 79 and its Central Compact Object  CXOU J1852$+$0040 \citep{seward03,halpern10}.  The spin period does not show any sign of Doppler shifts due to a possible companion star (and no companion star is observed in the optical or infrared bands in the available catalogs; Zhou et al. 2013). Assuming the pulsar being isolated, its rotational properties indicate a dipolar surface magnetic field (at the equator) $B=3.2\times10^{19} (P\dot{P})^{1/2} < 4\times10^{13}$\,G, the characteristic age and the rotation power are $\tau_c=P/(2\dot{P}) > 1.3$ Myr and $\dot{E}_{rot}=3.9\times10^{46}\dot{P}/P^{-3} < 3.5\times10^{30}$\ergs, respectively. Despite the relatively low dipolar magnetic field, the detection of an outburst and the observed spectral characteristics confirm the magnetar nature of this transient source.

In Figure~\ref{fig:mt} we show the expected timing and luminosity evolution for isolated neutron stars born with a dipolar field intensity (at the pole) between $B=10^{14}$\,G and $B=10^{15}$\,G. The expected properties are in line with the observed properties of all high-B pulsars, X-ray emitting Isolated Neutron Stars (XINS) and magnetars\footnote{Note that the high luminosity of some young magnetars is likely to be partly due to the contribute of the magnetospheric plasma, which yields part of its kinetic energy to X-ray photons via RCS. Thus, it is hard to infer the purely thermal component in the X-ray spectrum; see \cite{vigano13} for details.}.
As the right panel shows, \3xmm\ is compatible with the expected evolution of a neutron star born with an initial dipolar magnetic field (at the pole) of $B\sim 3-4\times 10^{14}$ G, which is now at the same evolutionary stage as the other low-B magnetars, hence with about a Myr. In particular, except for the occasional outburst, the long spin period $P\sim11.57$\,s  and the relatively low quiescent luminosity $L_{qui} < 4 \times10^{33}$\ergs of \3xmm\, (and of the other low-B magnetars \lowba\, and \lowbb\ ; \citealt{rea12,scholz12,rea13}) would place it in the same class as the XINS, a group of nearby, thermally emitting isolated neutron stars with typical temperatures of 0.1\,keV. The possibility that XINS or some of the other high-B pulsars are simply aged, less active magnetars has been proposed and tested in the last years (see e.g. \citealt{pons11,vigano13, rea13} and references therein). 

In this scenario, \3xmm lies on the same evolutionary track as the large group of other 6-7 neutron stars (magnetars and XINSs), perhaps indicating a common, typical initial magnetic field for these magnetars. The upper limit of $\dot{P}$ can be translated, within our evolutionary model, 
into a lower limit on the age of $\gtrsim 100$ kyr. In our theoretical models sporadical outbursts are expected to occur until a maximum age of $\sim 1$ Myr, after which the magnetic field is too weak to cause any significant crustal fracture. Thus, we can roughly estimate the real age between $\sim 0.1-1$ Myr (see also Figure~\ref{fig:mt} left panel). 
The upper limit we obtained for the quiescence luminosity is also compatible with the theoretical expectations for that age of $\sim 10^{32}-10^{33}$ erg~s$^{-1}$ (see left panel of Figure~\ref{fig:mt}).
\3xmm\, has the second largest period among isolated X-ray neutron stars, after 1E 1841--045, which confirms the clustering of periods of magnetars and XINS in a narrow range, not exceeding $\sim 12$\,s, and reinforces the idea that there must be a physical mechanism limiting the spin period \citep{pvr13}. 

In summary, all the outburst characteristics of \3xmm\, are typical of magnetars \citep{rea11}, with the outburst decay compatible with the crustal cooling scenario. This discovery supports the scenario in which magnetar-like activity is also expected in {\it normal} neutron stars, with inferred dipolar fields lower than the typical magnetar strength $10^{14}-10^{15}$\,G. A stronger crustal/internal field can be the responsible of the bursting activity of these aged magnetars \citep{rea10,turolla11,rea13}, although 
with a much lower event rate (less than $\sim$ one outburst per millennium; see \citealt{perna11}) than younger objects.  A simple estimate (see the discussion in \S 8.2 of \citealt{rea13}) gives an expected outburst rate for the entire population of low B magnetars in the galaxy of $\approx$ 1 per year. This number has to be confirmed by more detailed population synthesis studies including possible observational biases and selection effects, but it would not be surprising that more and more of these events are observed (or found in archival data after a more careful revision) in the upcoming years. 

Our last remark is about the possible association to the SNR Kes 79. Despite their apparent vicinity and a similar value of $N_{\rm H}$, this SNR is estimated to be $\sim 5-7$ kyr-old \citep{sun04}, much younger than the magnetar, and it hosts the CCO J1852$+$0040.  We then find unlikely a possible association between the magnetar and Kes 79.

\acknowledgments
NR is supported by a Ram\'on~y~Cajal fellowship and by an NWO Vidi Award. NR, DV and DFT acknowledge support by grants AYA2012-39303, SGR2009-811 and iLINK 2011-0303. JAP acknowledges support from the grants AYA 2010-21097-C03-02 and Prometeo/2009/103.  We thank Paolo Esposito for suggestions about the usage of the {\tt sosta} Ximage tool, Sergei Popov for pointing out a mis-reference in a draft version of this paper, and the referee for useful suggestions.

\label{lastpage}

\end{document}